\documentclass[10pt,a4paper]{article}

\usepackage[dvips]{graphicx}
\usepackage{fancyhdr}
\usepackage{booktabs}
\usepackage{amsmath}
\usepackage{amssymb}
\usepackage{lscape}
\usepackage{bm}
\usepackage{float}
\usepackage{comment}
\usepackage{url}
\usepackage{hyperref}
\usepackage{calc}
\usepackage[top=30truemm,bottom=30truemm,left=28truemm,right=28truemm]{geometry}
\usepackage{setspace}
\usepackage{cite}
\usepackage{upgreek}
\usepackage{indentfirst}

\usepackage{caption}
\usepackage{titlesec}
\titleformat*{\section}{\normalsize\bfseries}
\titleformat*{\subsection}{\normalsize\itshape}

\let\OLDthebibliography\thebibliography
\renewcommand\thebibliography[1]{
  \OLDthebibliography{#1}
  \setlength{\parskip}{0pt}
  \setlength{\itemsep}{3pt plus 0.3ex}
}

\begin{document}

$\,$

\vspace{60pt}

\begin{spacing}{1.8}
\begin{flushleft}
{\bf\LARGE Probing Physics in Vacuum Using an X-ray}

{\bf\LARGE Free-Electron Laser, a High-Power Laser,}

{\bf\LARGE and a High-Field Magnet}
\end{flushleft}
\end{spacing}

\vspace{-10pt}

\begin{flushright}
\begin{minipage}{0.85\hsize}
\begin{spacing}{1.2}
{\bf T. Inada$^{\,1,}$*, T. Yamazaki$^{\,1,}$*, T. Yamaji$^{\,2}$, Y. Seino$^{\,2}$, X. Fan$^{\,2}$,\\S. Kamioka$^{\,2}$, T. Namba$^{\,1}$ and S. Asai$^{\,2}$}
\end{spacing}

\vspace{5pt}

{\small
$^{1}$ \quad International Center for Elementary Particle Physics,

$\, \:$ \quad The University of Tokyo, 7-3-1 Hongo, Bunkyo-ku, Tokyo 113-0033, Japan\\
$^{2}$
\quad Department of Physics, Graduate School of Science,

$\, \:$ \quad The University of Tokyo, 7-3-1 Hongo, Bunkyo-ku, Tokyo 113-0033, Japan
}

\vspace{5pt}

E-mail: {\tt tinada@icepp.s.u-tokyo.ac.jp, yamazaki@icepp.s.u-tokyo.ac.jp}

\vspace{15pt}

{\bf Abstract:} A nonlinear interaction between photons is observed in a process that involves charge sources.
To observe this process in a vacuum, there are a growing number of theoretical and experimental studies.
This process may contain exotic contribution from new physics beyond the Standard Model of particle physics, and~is probed by experiments using a high-power laser or a high-field magnet, and more recently using an X-ray Free-Electron Laser (XFEL).
Here, we review the present status of our experiments testing various vacuum processes.
We describe four experiments with a focus on those using an XFEL: (i) photon--photon scattering in the x-ray region, (ii) laser-induced birefringence and diffraction of x rays, (iii) vacuum birefringence induced by a high-field magnet, and (iv) a dedicated search for axion-like particles using the magnet and x rays.
\end{minipage}
\end{flushright}

\vspace{8pt}

\hrule

\vspace{12pt}

\section{Introduction}\label{sec:introduction}

\begin{figure}[!t]
\begin{minipage}{0.5\hsize}
\centering
\includegraphics[clip,width=45mm]{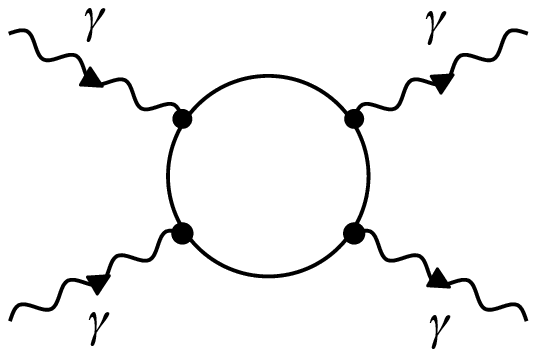}
\end{minipage}
\begin{minipage}{0.5\hsize}
\centering
\includegraphics[clip,width=45mm]{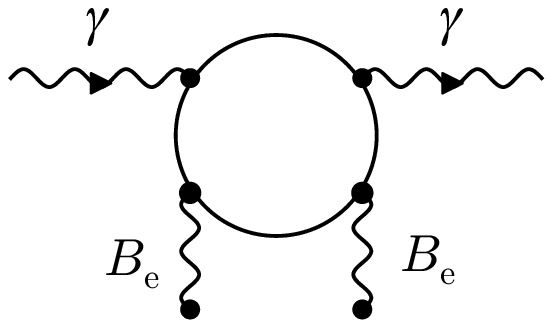}
\end{minipage}
\caption{
Vacuum processes of photon--photon interaction at the lowest order QED.
The time goes from left to right.
(Left)
Photon--photon scattering.
(Right)
Interaction of a photon with an external magnetic field that causes vacuum magnetic birefringence.
A virtual photon that conveys an electric or magnetic field is shown by a vertical line.
\label{fig:vacuum}}
\end{figure}

\begin{figure}[!t]
\begin{minipage}{0.24\hsize}
\centering
\includegraphics[clip,width=35mm]{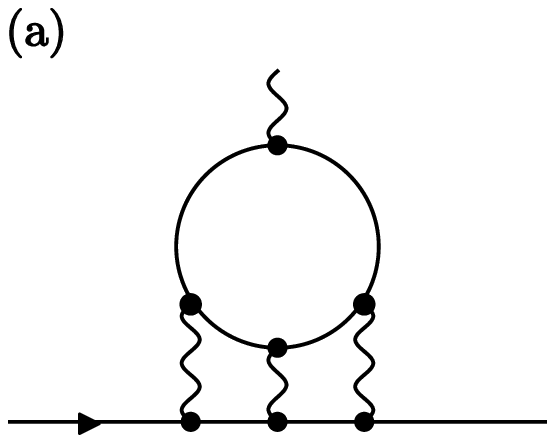}
\end{minipage}
\begin{minipage}{0.24\hsize}
\centering
\includegraphics[clip,width=35mm]{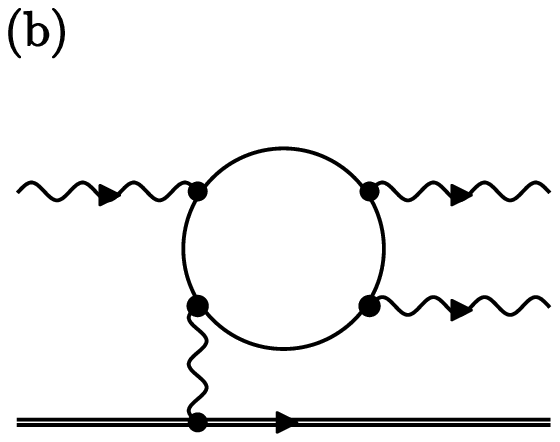}
\end{minipage}
\begin{minipage}{0.24\hsize}
\centering
\includegraphics[clip,width=35mm]{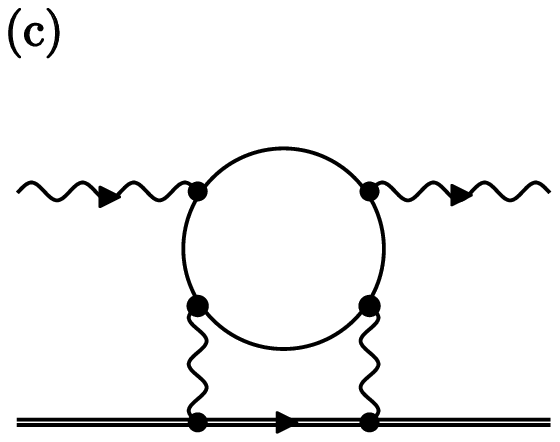}
\end{minipage}
\begin{minipage}{0.24\hsize}
\centering
\includegraphics[clip,width=35mm]{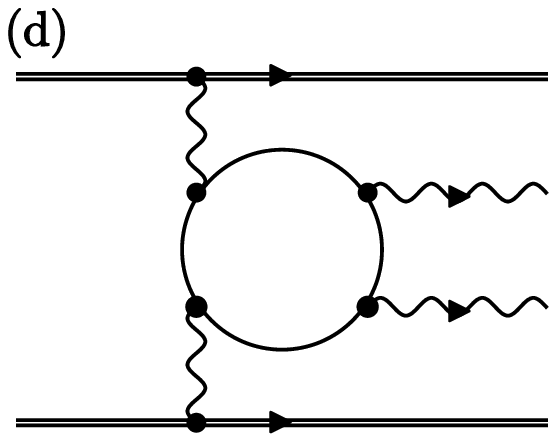}
\end{minipage}
\caption{
Real-charge processes containing a one-loop photon--photon interaction (see also Ref.~\cite{schlenvoigt} and the references therein).
(a) Radiative correction to the electron $g-2$ at the third order on the fine structure constant, with all relevant photons virtual~\cite{g-2_el,g-2_el_mea}.
(b) Photon splitting with one virtual photon from a nuclear field~\cite{ps}.
(c) Delbr\"uck scattering with two virtual photons~\cite{ds}.
(d)~Pair production from the crossing of two nuclear fields~\cite{upc}.
\label{fig:real_charge}}
\end{figure}

Photon--photon scattering is a nonlinear interaction between photons, intermediated by a virtual electron loop at the lowest order of quantum electrodynamics (QED) (Fig.~\ref{fig:vacuum}, left).
The~contribution of this diagram ubiquitously appears in radiative corrections to charged particles at higher orders~(Fig.~\ref{fig:real_charge}a),
and its agreement with measurements constitutes a solid basis for the Standard Model of particle physics at low energies~\cite{g-2_el, g-2_mu, g-2_el_mea}.
In addition, some of these photons have been changed into real-state photons (Fig.~\ref{fig:real_charge}b--d) and observed in experiments~\cite{ps, ds, upc}.
These processes occur under a Coulomb field of pre-existing, real-state charges, and therefore can be classified as ``real-charge~processes''.
By contrast, there are many attempts to observe this diagram in a vacuum by using a PW-class high-power laser or using a high-field magnet.
This kind of ``vacuum process'' occurs without such charge sources and cannot be removed by any effort, thus representing the {\it core} of all physical processes.

The vacuum process is currently studied in two streams.
The first is based in high field science, where the behavior of dense plasma is studied using high-power lasers.
They predict many kinds of vacuum phenomena occurring at the focal spot of an intense laser~\cite{review1, review2}.
An efficient probe on these phenomena is provided by the recent progress in X-ray Free-Electron Lasers (XFELs) due to their short wavelength, peak brilliance, and more particle nature compared to laser photons~\cite{schlenvoigt}.
The~second is an experimentally ongoing stream that studies the optical birefringence of a magnetized vacuum~(Fig.~\ref{fig:vacuum}, right).
The measurement of vacuum magnetic birefringence (VMB) provides the current best sensitivity to the observation of the vacuum process~\cite{pvlas}.

The theoretical basis describing these vacuum processes can be formulated in terms of the Euler--Heisenberg Lagrangian, which effectively contains only electromagnetic fields.
With~assumptions that photon energies are well below the electron mass $m_{{\rm e}}$ and fields much smaller than the QED critical values ($E \ll 1.3 \times 10^{18}$ ${\rm V\, m^{-1}}$, $B \ll 4.4 \times 10^{9}$ ${\rm T}$), the Lagrangian density is written in natural units~($\hbar = c = 1$) as~\cite{eh}
\begin{equation}
\mathcal{L} = \frac{1}{2} \left( E^{2} - B^{2} \right) + \frac{2\alpha^{2}}{45m_{\rm{e}}^{4}} \left[ \left( E^{2} - B^{2} \right) ^{2} + 7 \left( \bf{E \cdot B} \right) ^{2} \right],
\label{eq:1}
\end{equation}
with the fine-structure constant $\alpha$ and electric and magnetic fields $\bf{E}$ and $\bf{B}$, respectively.
The second term of Eq.~(\ref{eq:1}) describes four-field interactions, and theoretical studies on these effects have been summarized in Ref.~\cite{review1, review2}.
In addition to these nonlinear QED contexts, searches for vacuum processes concern a wider range of research fields, including astrophysics and particle physics.
For instance, a practical effect of VMB has been suggested by the recent observation of a strongly magnetized neutron star~\cite{neutron_star1, neutron_star2, neutron_star3}.
Furthermore, the vacuum process may be easily affected by the contribution of new physics beyond the Standard Model (BSM)~\cite{snow_mass}.
The conventional benchmark in a low energy scale is the axion, and more generally, axion-like particles (ALPs) that have an effective coupling to two photons~\cite{axion_review}.
If such light particles exist, their mass spectrum would increase the diversity of vacuum interactions ascribed to electrons at low energies.

To our best knowledge, we are currently the only group that experimentally studies the vacuum physics in both streams.
In this article, we review the status and results of our first-phase experiments along with their technical aspects.
Section~\ref{sec:optics} describes experiments probing the vacuum~process; two~experiments using the XFEL, SACLA (SPring-8 Angstrom Compact Free-Electron Laser)~\mbox{\cite{sacla1, sacla2}}, and one for VMB using a pulsed magnet.
BSM searches by these experiments are discussed in Sec.~\ref{sec:bsm}, particularly in the case of ALPs.
A dedicated search using the pulsed magnet and XFEL is also~described.

\begin{figure}[!t]
\centering
\includegraphics[clip,width=120mm]{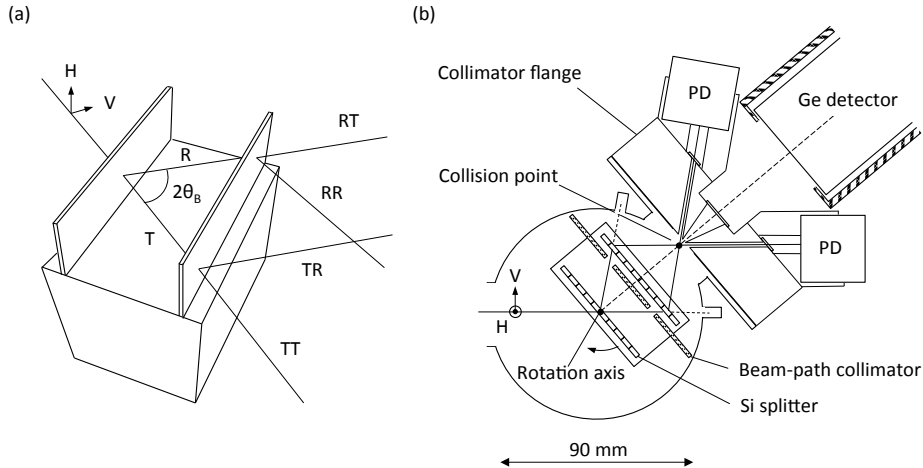}
\caption{
(a) Four beams after the two blades of the silicon beam splitter.
(b) Magnified view of the beam-collision chamber.
Adapted from Ref.~\cite{gg1}.
\label{fig:gg_setup}}
\end{figure}   

\section{Probing the Vacuum Process}\label{sec:optics}

So far, no experiments have observed the vacuum process due to its extremely weak signal.
Signal~photons are characterized by (i) the frequency, (ii) the wave vector, and (iii) the polarization.
Any~vacuum process is finally detected as the changes in these observables.
This simple fact results in a variety of macroscopic phenomena probed in experiments~\cite{review1, review2}.

\subsection{Elastic Scattering by the Collision of Two X-ray Pulses}\label{subsec:gg}

This section reviews our experiments at SACLA that provide the first limits on the QED cross-section of photon--photon scattering ($\sigma_{\gamma\gamma}$) in the x-ray region~\cite{gg1,gg2}.
Before the experiments, the scattering was tested by using optical lasers~\cite{moulin}.
However, the cross-section is suppressed by the sixth power of the ratio $\hbar \omega_/m_{{\rm e}}c^{2}$,
where $\omega$ is the photon energy and $m_{{\rm e}}$ the electron mass~\cite{gg_xsec1, gg_xsec2}.
Therefore, a significant enhancement can be obtained by using high energy x-ray photons.

It is not an easy task to collide x-ray pulses if they are delivered as two distinct beams.
Thus,~we~developed a new scheme to divide a single beam into two and then to collide them.
This~scheme applies a technology established for an x-ray interferometer.
A schematic of the beam splitter is shown in Fig.~\ref{fig:gg_setup}a.
A beam is split into two coherent beams---one is the transmitted beam (T) and the other is the reflected one (R)---by a 0.6~mm-thick blade of Si(440) crystal when the lattice plane matches a diffraction angle of $\theta_{{\rm B}}=36^\circ$~\cite{bonse1}.
A similar process occurs at the second blade, creating four beams after the two blades.
The TR and RR beams collide obliquely with a crossing angle of $2 \theta_{{\rm B}}$, and the two-photon system is boosted along $\theta_{{\rm B}}$.
The change of wave vectors in the center-of-mass system causes energy unbalance of the pairs observed in the laboratory frame.
This is an example of a kinematically-induced frequency change occurring in oblique collisions, in contrast to one caused by optical nonlinearities such as four-wave mixing~\cite{bernard}.
In this way, the diffraction condition itself ensures the geometrical equivalence of the beam paths split by the symmetric Laue reflection~\cite{bonse2}.

\begin{figure}[!t]
\begin{minipage}{0.5\hsize}
\centering
\includegraphics[clip,height=57mm]{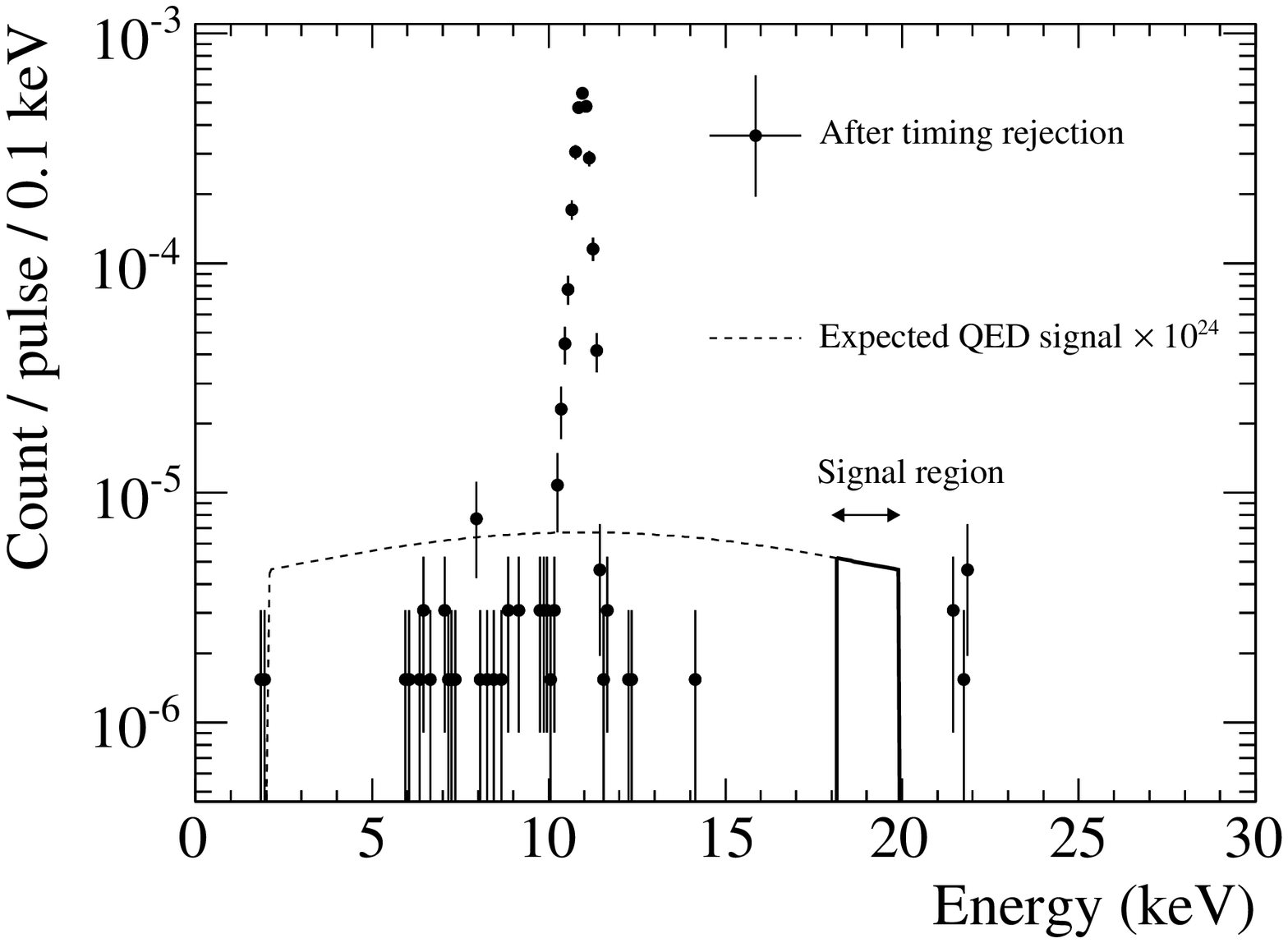}
\end{minipage}
\begin{minipage}{0.5\hsize}
\vspace{2mm}
\centering
\includegraphics[clip,height=57mm]{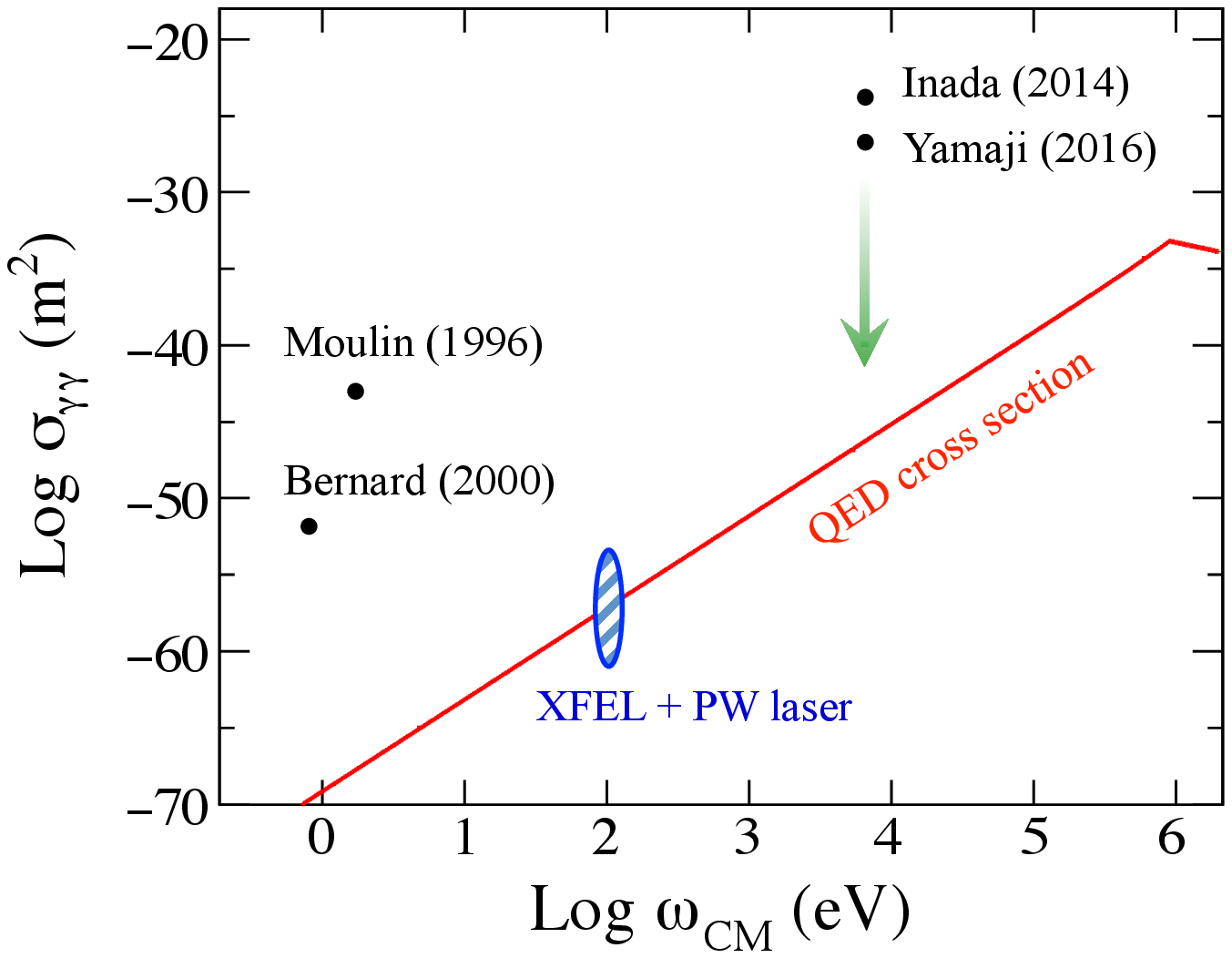}
\end{minipage}
\caption{
(Left)
Energy spectrum obtained from the total injection of $6.5\times10^{5}$ pulses, after~the timing rejection cut.
The dotted line shows the expected QED signal shape in the full solid angle~(signal strength is $\times10^{24}$).
The solid line corresponds to the signal region from 18.1~keV~($\theta=12.5^{\circ}$) to 19.9~keV ($\theta=0$).
Adapted from Ref.~\cite{gg1}.
(Right)
Limits on $\sigma_{\gamma\gamma}$ at 95\% C.L. obtained by real photon--photon scattering experiments.
Moulin~(1996)~\cite{moulin} uses two high-power~lasers.
Bernard~(2000)~\cite{bernard} uses three lasers applying four-wave mixing.
Inada~(2014)~\cite{gg1} is our first search using an XFEL, and we upgraded it in Yamaji~(2016)~\cite{gg2}.
The hatched circle shows a region probed by laser-induced diffraction and birefringence experiments using an XFEL.
\label{fig:gg_result}}
\end{figure}   

A magnified view of the beam-collision chamber is shown in Fig.~\ref{fig:gg_setup}(b).
The splitter is installed in the chamber and fixed to a rotary feedthrough connected to a goniometer.
The intensities of the two beams are monitored pulse-by-pulse by silicon PIN photodiodes (PDs).
One of the paired signal~photons---the forward-scattered one---has higher energy than the original photons~($\omega_{0}=11$~keV), and is easily separated from backgrounds.
To select forward-scattered photons, a collimator flange with a 
conically tapered hole is placed after the collision point.
The full angle of the cone is~25$^{\circ}$, defining the signal energy region from 18.1~keV to 19.9~keV.
The first experiment was performed on 23rd and 24th July 2013 at SACLA with a total injection of $6.5\times10^{5}$ pulses.
A timing cut was applied to select events within $\pm 1$ $\upmu$s of the arrival of x-ray pulses.
Figure~\ref{fig:gg_result}(left) shows the energy spectrum after the cut.
An elastic peak from stray x rays is observed around $\omega_{0}$ as well as some events at 2$\omega_{0}$ due to the pileup of two photons.
No events are observed in the signal region, and the resulting limit on $\sigma_{\gamma \gamma}$ is shown in Fig.~\ref{fig:gg_result}(right) with other limits.

After the first experiment, we upgraded it in the second experiment carried out in 2015~\cite{gg2}.
The~main improvements are in (i) the efficiency of the beam splitter, (ii) the performance of the~XFEL, and (iii) the statistical gain by larger measurement time.
The total improvement from the first result is by three orders of magnitude, and gives the most stringent limit in the x-ray~region, which still needs to be gained by 20 orders of magnitude to reach the QED cross-section.
On the basis of these experiences, we are planning to further improve the sensitivity in the next~phase.
The~first approach is to apply a similar splitter in the Bragg case as seen in Ref.~\cite{bonse3}.
This reduces the absorption and beam-branching in the second blade of the Laue case used in the present~splitter.
Moreover, multiple collisions are possible by bouncing two x-ray pulses many times in the same~crystal.
The second gain would be obtained by improving the performance of the XFEL.
The present mechanism of the x-ray laser is called ``self-amplified spontaneous emission (SASE)'', where the output beam does not have deterministic monochromaticity, reflecting the stochastic nature of initial electron~bunches.
An alternative ``self-seeding'' scheme has been proposed, and is expected to improve the current monochromaticity by about two orders of magnitude, reducing the monochromaticity loss in the splitter~\cite{seed}.
Finally, the direct collision of distinct beams, with each tightly focused at a collision~point, can significantly improve the sensitivity.
There is a dedicated facility between SACLA and SPring-8 where both beams are available~\cite{double}.
The temporal synchronization of the two beams is currently underway.

\subsection{Laser-Induced Birefringence and Diffraction of X rays}\label{subsec:vd}

\begin{figure}[t]
\begin{minipage}{0.5\hsize}
\centering
\includegraphics[clip,width=65mm]{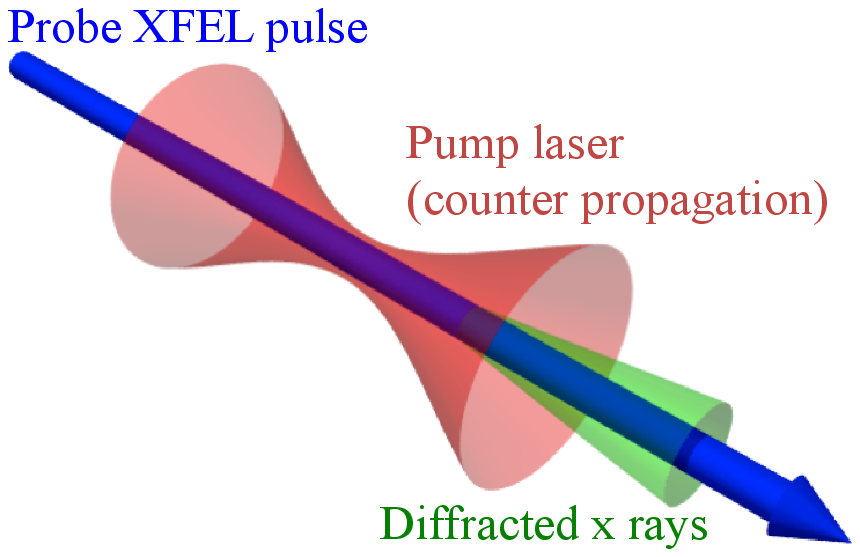}
\end{minipage}
\begin{minipage}{0.5\hsize}
\centering
\includegraphics[clip,width=60mm]{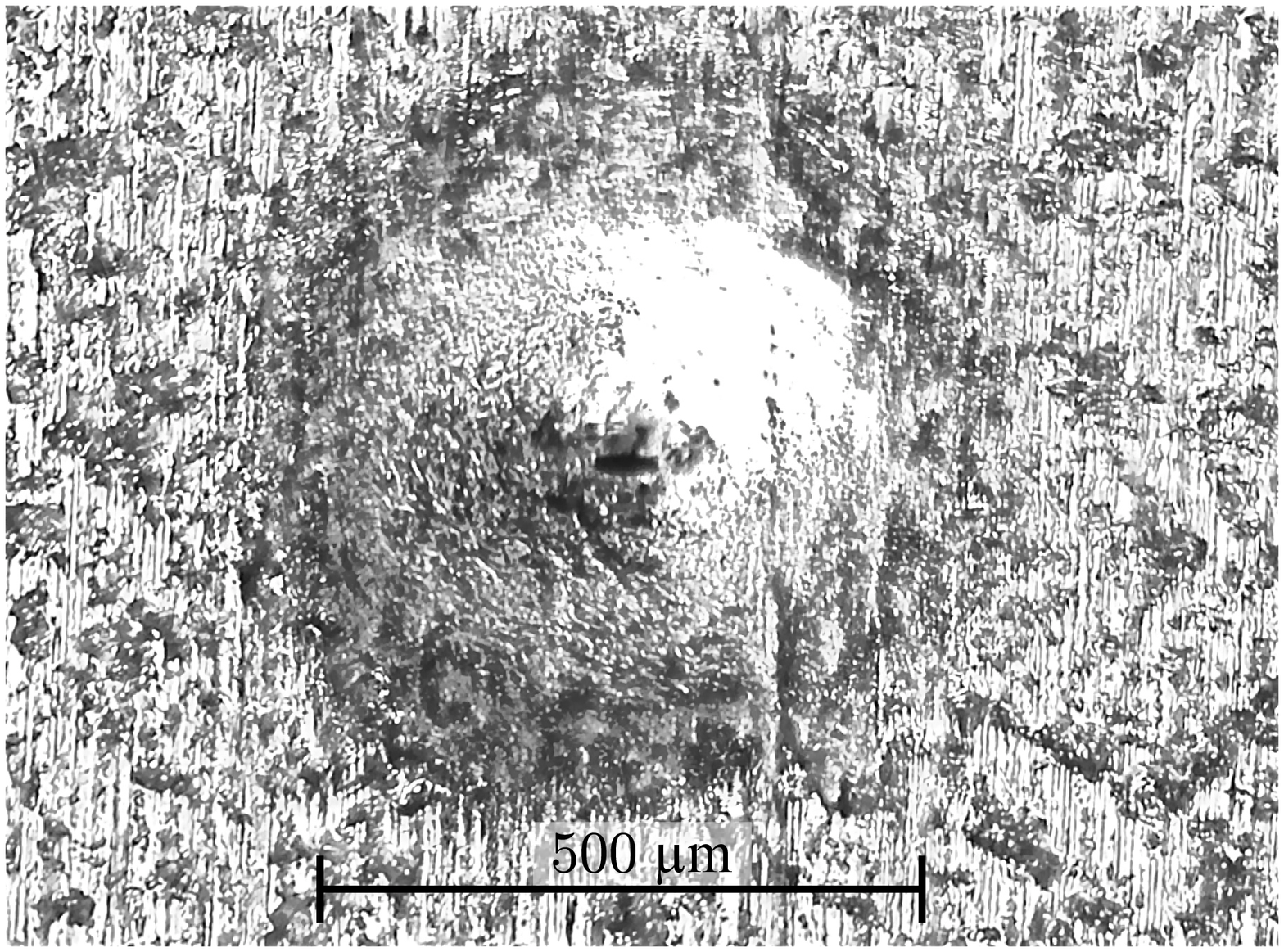}
\end{minipage}
\caption{
(Right)
Principle of the experiment probing vacuum diffraction.
(Left)
Microscope image of a zinc film after the temporal and spatial adjustments.
The rectangular hole at the center is made by the penetration of x rays (40(H) $\times$ 20(V) $\upmu$m).
The large deformation overlapping the hole is made by a focused 2.5-TW laser (12(H) $\times$ 10(V) $\upmu$m).
\label{fig:vd}}
\end{figure}   

An XFEL can work as a useful probe on the vacuum process induced at the focal spot of a high-power laser.
Since both lasers provide real photons, this is another real process corresponding to a different center-of-mass photon energy $\sqrt{\omega_{1} \omega_{2}}=100$~eV with $\omega_{1}=10$~keV and $\omega_{2}=1$~eV .
The high-power laser is often referred to as a pump laser in analogy to a conventional pump-probe experiment where the laser excites a sample and the probe studies the response.
The change of the probe's wave vector induced by the scattering is called ``vacuum diffraction''~\cite{vd}.
Furthermore, polarization flipping---which occasionally occurs in the scattering---adds tiny ellipticity to the linearly polarized probe, and is called ``vacuum birefringence''~\cite{vb}.
Figure~\ref{fig:vd}(left) shows the principle of the experiment where the pump laser is radiated in a counter-propagating geometry.
The probe and pump are focused on the same spot to enhance the diffraction signal that is scattered in a forward direction with a tiny angle.
An efficient scheme of detecting photons that have both properties of flipped polarization and a finite diffraction angle was proposed in Ref.~\cite{karbstein}.
In addition, a practical setup using an x-ray focusing lens and crossed polarizers can be found in Ref.~\cite{schlenvoigt}.
While there are many theoretical studies and proposals, no experiments have been put into practice.

Regardless of detecting diffraction or birefringence, the key point of the experiment is the way to synchronize the probe and pump with respect to space and time.
As for the timing adjustment, there is a working method using the fast electronic response of a GaAs film that can tune the timing within $\sim$1~ps~\cite{katayama}.
However, to our knowledge, there is so far no efficient method to spatially overlap the pump and probe with both focused to $\sim$10~$\upmu$m,
and thus a dedicated method needs to be found and established.

For this purpose, we carried out a test measurement at SACLA in 2016.
A probe XFEL and a 2.5-TW pump laser were focused to about 10~$\upmu$m by composite refractive lenses and an off-axis parabolic mirror, respectively.
A zinc film with a thickness of 20~$\upmu$m was placed at the collision point to check their overlap (the detailed setup and procedures are described in Ref.~\cite{seino}).
Their energy deposit on the film creates a characteristic pattern of deformation that could be checked later by an optical microscope.
Figure~\ref{fig:vd}(right) shows a microscope image of the film after the timing adjustment and some iteration of the spatial overlapping.
The XFEL was shot on the film from the back side of the image.
The larger hole with a diameter of about 500~$\upmu$m was made by the focused pump laser, whereas a smaller one at the center was made by the penetration of the probe.
The deviation between the two centers was found to be less than $\sim$10~$\upmu$m, showing the accuracy of this method~\cite{seino}.

A PW-laser facility is under installation at SACLA and is scheduled to be available to users in~2018~\cite{double}.
The expected sensitivity to $\sigma_{\gamma \gamma}$ is shown in Fig.~\ref{fig:gg_result}(right).
By the time the laser system becomes ready for use, our effort will be directed to the establishment of a synchronization scheme feasible for a higher pump power.

\subsection{Birefringence Induced by a High-Field Magnet}\label{sec:vmb}

Magnet-induced birefringence has long been measured in normal media like gas, and is known as the ``Cotton-Mouton effect''~\cite{cm}.
The application of this measurement scheme to a vacuum provides the current best sensitivity to observe the vacuum process.
The most recent result of the PVLAS experiment in 2016 provides sensitivity that is smaller by a factor of 20 with respect to the coefficient of the vacuum refractive index~\cite{pvlas}.
The signal birefringence increases in proportion to the square of the magnetic field strength, and thus a high field of a pulsed magnet is advantageous.
In addition, the temporal variation of the pulsed field is essential to distinguish the signal from large static birefringence produced by cavity mirrors.
In this scheme, a racetrack pulsed magnet was first used in the BMV experiment~\cite{bmv}.

\begin{figure}[!t]
\centering
\includegraphics[clip,width=75mm]{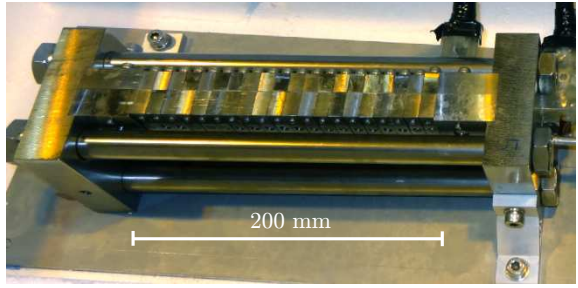}
\caption{
Racetrack pulsed magnet developed for a high-repetition use.
The field length is 200~mm.
\label{fig:mag}}
\end{figure}   

\begin{figure}[!t]
\begin{minipage}{0.5\hsize}
\centering
\includegraphics[clip,height=45mm]{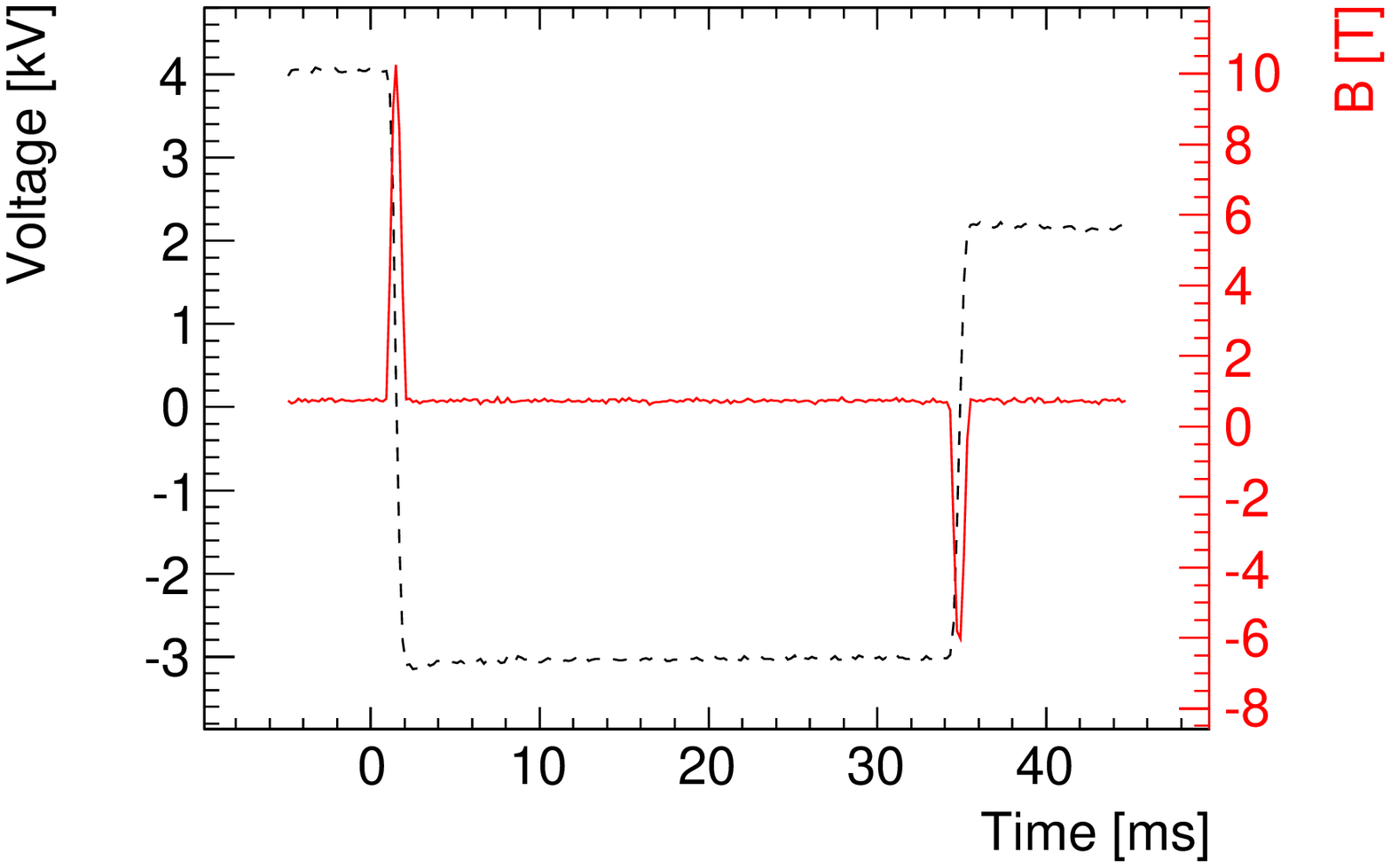}
\end{minipage}
\begin{minipage}{0.5\hsize}
\centering
\includegraphics[clip,height=45mm]{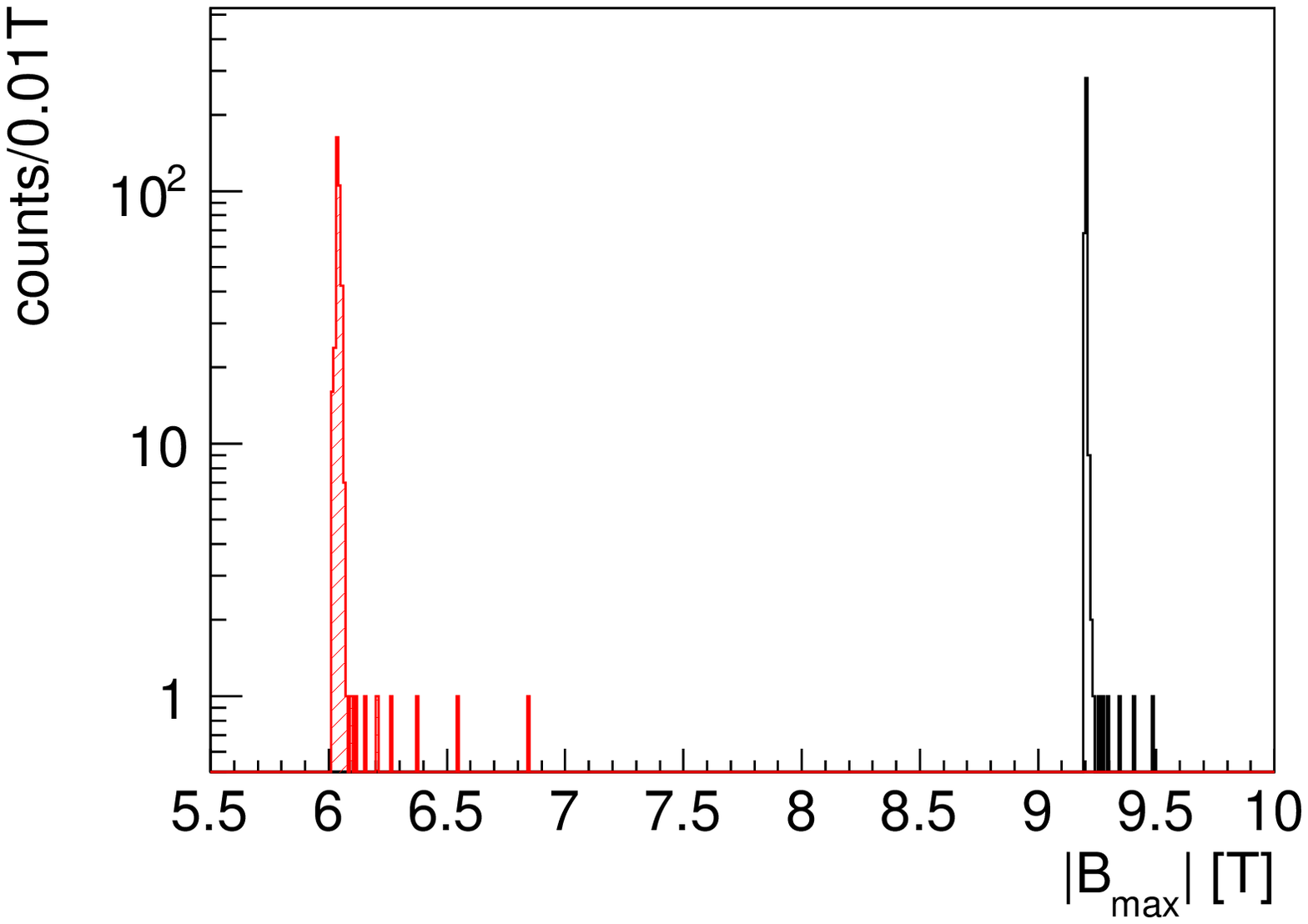}
\end{minipage}
\caption{
Repetitive operation of the four pulsed magnets.
(Left)
Typical waveform of the capacitor voltage (to left, dashed) and magnetic field (to right, solid).
The pulse interval is adjustable and set to $1/30$~s here.
(Right)
Peak-field distribution at the magnet center for the first and second~pulses, obtained by 1-h operation with 0.2-Hz pulse repetition.
Adapted from Ref.~\cite{pulse_mag}.
\label{fig:rep}}
\end{figure}   

To obtain large statistics along with the field strength, we have been developing a pulsed magnet system suited to a high repetition use.
While the typical repetition of a pulsed magnet is about 1~mHz, that of our magnet is higher by about two orders of magnitude~\cite{pulse_mag}.
Figure~\ref{fig:mag} shows a photo of the magnet.
The coil part is wound with a copper wire and cooled with liquid nitrogen to reduce the electrical resistivity.
Stainless steel metal is used to reinforce the outer strength of the coil that is exposed to an expansion force during the generation of pulsed fields.
The peak field reaches 12 T at the magnet center.
The magnet has a planar shape to enhance the cooling efficiency.
In addition to the magnet, we also developed a discharge unit with capacitors providing pulsed current to the magnet with high repetition~\cite{pulse_mag}.
Figure~\ref{fig:rep}(left) shows a cycle of the repetition composed of two successive pulses with alternating field directions.
The first pulse produces a field of $+10$ T at the magnet center with a charged voltage of 4 kV.
After a portion of the energy converts into Joule heating in the coil, the second pulse produces $-6$ T.
The residual energy returns to the capacitor that is recharged to 4 kV before the next cycle.
Figure~\ref{fig:rep}(right) shows the distribution of the field strength in a continuous operation.
Stable fields are repeatedly produced after the magnets relax to thermal equilibrium.
The field length linearly increases the signal birefringence.
Up to now, we have tested a simultaneous operation of four magnets, giving a total field length of 0.8 m with a repetition rate of 0.2 Hz and a field strength of 10 T (see also Sec.~\ref{subsec:alp} and Fig.~\ref{fig:lsw_setup}).

We have made a setup that combines the magnet with an optical system containing a Fabry--Perot cavity with a finesse of 650\,000.
The continuous data-taking requires a new scheme that the cavity resonance automatically recovers after the measurements of finesse and static birefringence in each~cycle.
To verify this scheme, we carried out a test run using a single magnet.
The results have been described in Ref.~\cite{fan}.
A long-term run of a few months will start after improvements of the~optics.

\section{Searches for Physics Beyond the Standard Model}\label{sec:bsm}

The subtle signal of the vacuum process would be easily affected by the possible existence of new~physics.
If a new particle couples to photons, it can be directly searched for by the same vacuum~experiment.
A good example was provided by the recent discovery of the Higgs boson which decays into two photons ($H \to \gamma \gamma$), demonstrating that a vacuum is permeated by a scalar~field~\mbox{\cite{higgs_atl,higgs_cms}.}
As lighter particles, new pseudoscalar bosons like ALPs provide anomalous contribution to the vacuum~process~\cite{axion_review}.
Moreover, BSM models often predict a dark sector that involves fermions with a tiny fraction of the electric charge, known as millicharged particles (see Ref.~\cite{snow_mass} for a systematic review of BSM in a low energy scale).
Among these particles, here we focus on the ALP case.

\subsection{Exotic Contribution to Photon--Photon Scattering}\label{subsec:exotic}

\begin{figure}[!t]
\centering
\includegraphics[clip,width=60mm]{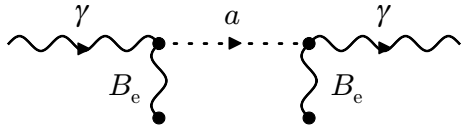}
\caption{
Interaction of a photon with an external magnetic field caused by an axion-like particle, contributing to vacuum magnetic birefringence.
\label{fig:alp_vmb}}
\end{figure}   

\begin{figure}[!t]
\begin{minipage}{0.5\hsize}
\centering
\includegraphics[clip,width=50mm]{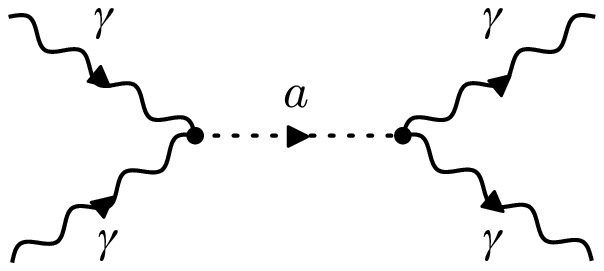}
\end{minipage}
\begin{minipage}{0.5\hsize}
\centering
\includegraphics[clip,width=50mm]{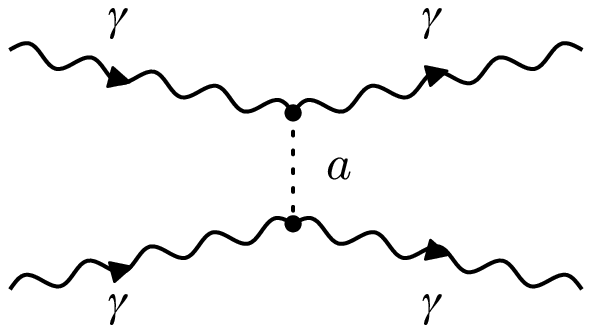}
\end{minipage}
\caption{
Photon--photon scattering by axion-like particles.
(Left) $s$-channel.
(Right) $t$-channel.
\label{fig:alp_gg}}
\end{figure}   

As stated before, the VMB measurement provides an efficient strategy for observing the vacuum~process, and this is also true for the ALP search for a mass region around 1~meV~\cite{pvlas}.
Figure~\ref{fig:alp_vmb} shows a diagram where ALPs contribute to VMB.
However, it inclusively searches for a wide mass region since the mass information of ALPs is somewhat lost by the virtual off-shell photons conveying magnetic~fields.
By contrast, the $s$-channel scattering of real photons provides mass information with a precision of the monochromaticity of the light [Fig.~\ref{fig:alp_gg}(a)].
Once the ALP mass is roughly determined, the $s$-channel search can precisely sweep the relevant mass region by continuously changing the incident photon energy.
This is advantageous for real photon--photon processes~\cite{gg1, gg2}, and thus the two methods provide complementary approaches.

Each kind of light source probes a distinct mass region corresponding to its photon energy.
As~shown in Fig.~\ref{fig:gg_result}(right), we are currently capable of the combinations of XFEL--XFEL, XFEL--laser, and laser--laser, for which the center-of-mass energies are about 10~keV, 100~eV and 1~eV, respectively.
Various combinations of these different light sources widen the experimental opportunities, since the new mass scale is still not clear.

\subsection{Dedicated Search for Axion-Like Particles}\label{subsec:alp}

\begin{figure}[!t]
\centering
\includegraphics[clip,angle=90,width=100mm]{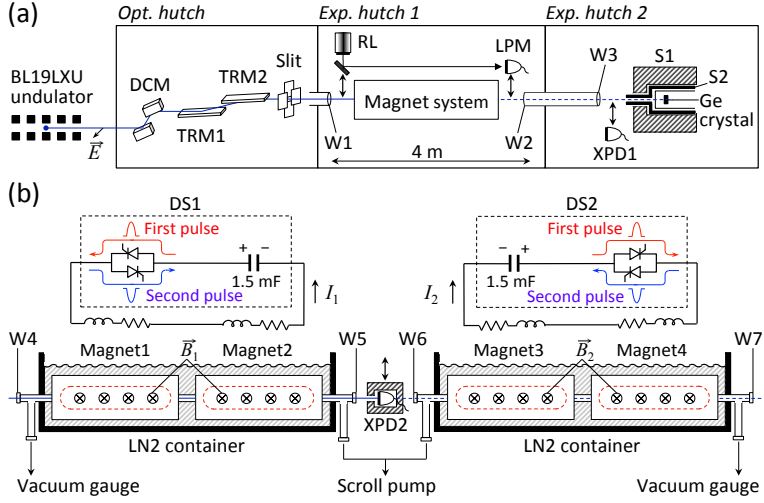}
\caption{
Schematic of the experimental setup.
(a) Layout of components in the optics hutch and in the two experimental hutches.
Retractable components are shown with a vertical arrow.
(b) Side view of the magnet system.
The magnets are placed so as to produce parallel fields with respect to the horizontal polarization of the x-ray beam.
Two identical discharge sections~(DS1 and~DS2) that are contained in the same hutch supply pulsed currents to the magnets by $LC$-discharge~circuits.
Adapted from Ref.~\cite{spring-8}.
\label{fig:lsw_setup}}
\end{figure}   

Compared to the mass region probed by VMB measurements, a heavier region can be studied by using x rays~\cite{paraphoton,spring-8}.
Especially, the ALP mass around 10--100~meV attracts recent cosmological interests in the inflation, the cosmic microwave background radiation, and dark matter~\cite{takahashi}.
The~photon conversion into a real-state ALP (or vice versa) in a magnetic field is referred to as the ``Primakoff effect''.
Up to now, a large number of laboratory searches for ALPs have been carried out with a ``light shining through a wall (LSW)'' technique~\cite{axion_review},
The first x-ray LSW search was carried out at the European Synchrotron Radiation Facility (ESRF) with superconducting magnets, extending~the limit on the ALP--two-photon coupling constant ($g_{a\gamma\gamma}$) up to around 1~eV~\cite{esrf}.
While~searches for ALP flux from the Sun have also probed this mass region~\cite{cast},
the flux estimation inevitably relies on a solar model and its complex magnetic activity~\cite{solar_mag}, showing the importance of complementary searches using terrestrial and extra-terrestrial x-ray sources~\cite{esrf}.

\begin{figure}[!t]
\begin{minipage}{0.5\hsize}
\centering
\includegraphics[clip,height=52mm]{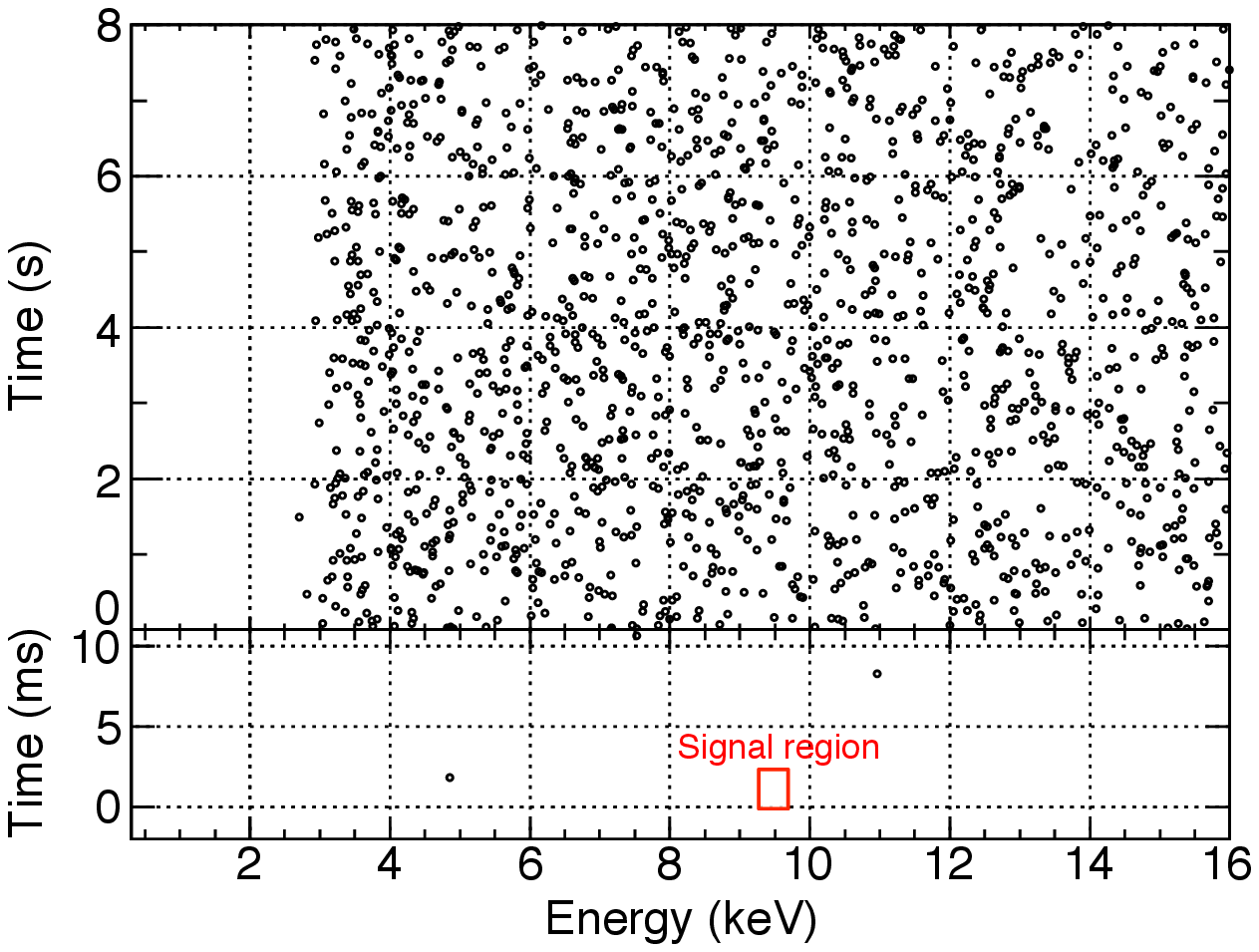}
\end{minipage}
\begin{minipage}{0.5\hsize}
\vspace{2mm}
\centering
\includegraphics[clip,height=54mm]{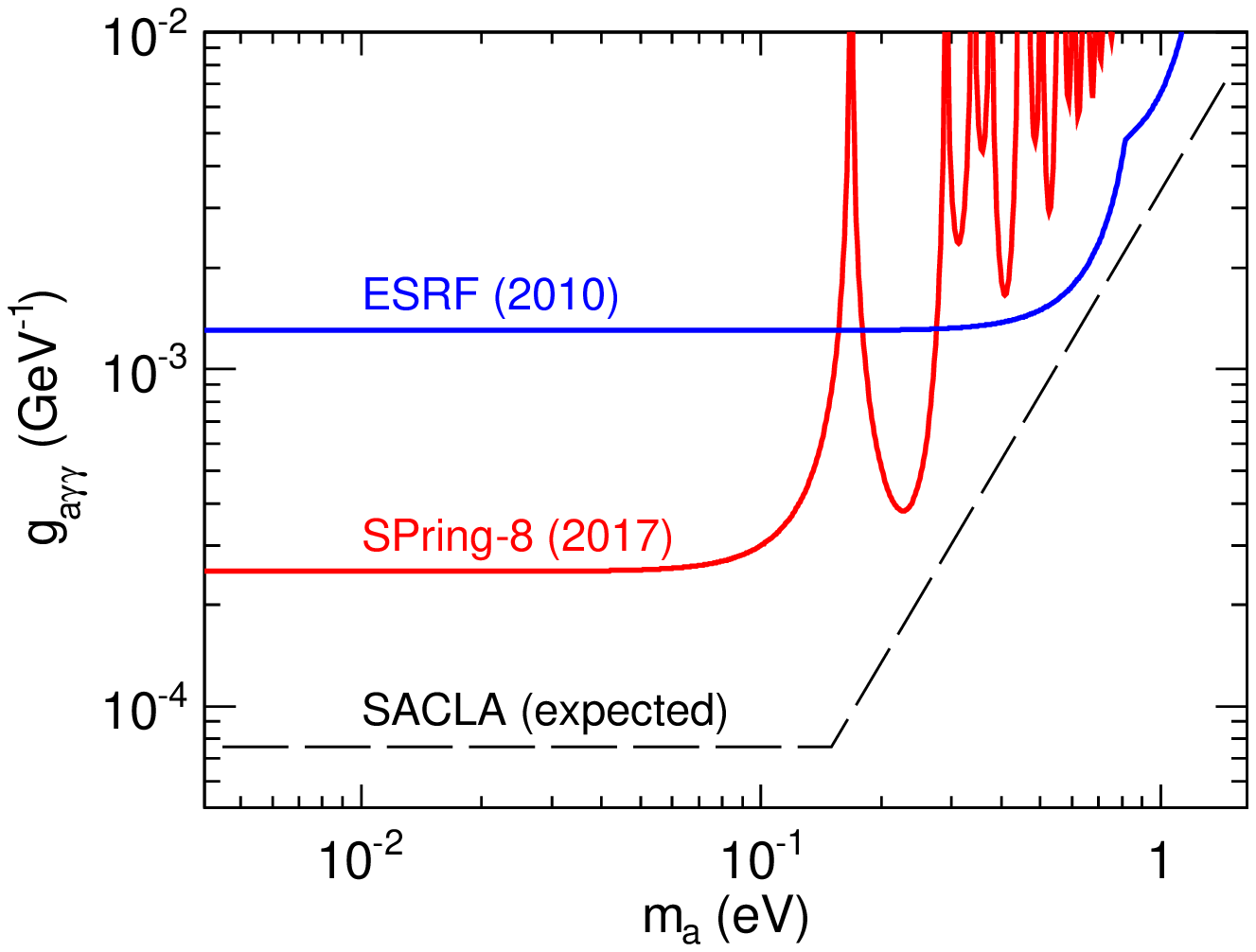}
\end{minipage}
\caption{
(Left)
Top:
time--energy distribution of x rays measured with the germanium detector.
Each circle represents an event.
The horizontal axis is the time from the beginning of the first~pulse.
The energy threshold was set around 3~keV.
Bottom: 
events around the signal region defined by the beam energy of $\pm 2 \sigma$ ($\sigma$ is the detector resolution) and the time window of 2.1 ms.
Adapted~from Ref.~\cite{spring-8}.
(Right)
Upper limit on the ALP--two-photon coupling constant $g_{a\gamma\gamma}$ at 95\% C.L. as a function of the ALP mass.
The x-ray LSW limits obtained at the ESRF~(2010)~\cite{esrf} and SPring-8~(2017)~\cite{spring-8} are shown along with the expected sensitivity at SACLA (dashed).
\label{fig:lsw_result}}
\end{figure}

Figure~\ref{fig:lsw_setup} shows a schematic of our LSW setup using four magnets described in Sec.~\ref{sec:vmb}.
The ALPs generated in the first pair of magnets pass through a beam dump that blocks the unconverted~photons.
Some of the ALPs then reconvert into detectable photons by an inverse process in the second magnet pair.
We carried out the first measurement at SPring-8 BL19LXU~\cite{bl19} in 2015 using a continuous x-ray beam with high intensity.
The data obtained with a total of 27\,676 pulses produced in a net run time of two days show no events in the signal region (Fig.~\ref{fig:lsw_result}, left).
This~provides a limit on $g_{a \gamma \gamma}$ that is more stringent by a factor of 5.2 compared to the previous x-ray LSW limit for the ALP mass $\lesssim0.1$~eV (Fig.~\ref{fig:lsw_result}, right).
The pulsed nature of the magnet provides better sensitivity when used with pulsed x rays (the repetition is 30 Hz at SACLA; see Fig.~\ref{fig:rep}(left) for the corresponding time interval between the magnetic pulses).
The expected sensitivity at SACLA with the same magnets and configurations are also shown in the figure.

\section{Summary}\label{sec:conclusion}

We reviewed the progress of our experiments that probe physics in a vacuum, with a focus on those using an XFEL.
Various combinations among an XFEL, a high-power laser, and a high-field magnet provided a test for QED and new physics in a different energy scale.
The results obtained by our first-phase experiments are as follows.
(i) The photon--photon scattering experiment using an XFEL provides the first limits in the x-ray region~\cite{gg1, gg2}.
The sensitivity needs to be gained by 20 orders of magnitude to reach the QED cross-section and now proceeds to the second phase with significant improvements.
(ii) The experiment probing laser-induced birefringence and diffraction of x rays requires a new synchronization scheme between the pump and probe lasers.
A test measurement using a 2.5-TW pump laser succeeded in the temporal and spatial synchronization with the accuracies of $\sim$1~ps and 10~$\upmu$m, respectively.
We are currently preparing for the first measurement using a PW laser in 2018~\cite{seino}.
(iii) The VMB measurement features a unique field-generation system with high-repetition pulsed magnets.
The use of a high-finesse Fabry-Perot cavity verified the automatic resonance recovery scheme required for continuous data-taking and obtained the first results~\cite{pulse_mag,fan}.
A long-term run for a few months will start after some improvements of the optics.
(iv) The dedicated search for ALPs using the magnets and continuous x rays was carried out at SPring-8 and improved the previous limit on the coupling constant by a factor of 5.2~\cite{spring-8}.
The preparation for an experiment at SACLA using pulsed x rays is underway.

\vspace{6pt} 

\section*{Acknowledgments}

This research work at SACLA and SPring-8 BL19LXU is approved and supported by JASRI and~RIKEN.

\end{document}